\documentclass{article}
\begin{document}
\title{On the Equivalence Principle of Quantum Mechanics}
\author{Jos\'e M. Isidro \\ 
Dipartimento di Fisica ``G. Galilei''\\ 
Istituto Nazionale di Fisica Nucleare\\ 
Universit\`a di Padova, Via Marzolo, 8\\ 
35131 Padova, Italy\\
{\tt isidro@pd.infn.it}}
\maketitle

\begin{abstract}
A recent concept in theoretical physics, motivated in string duality and 
M--theory, is the notion that not all quantum theories arise from quantising 
a classical system. Also, a given quantum model may possess more than just one 
classical limit. In view of these developments, we analyse some general properties 
that quantum mechanics must satisfy, if it is {\it not} to be formulated as a quantisation 
of a given classical mechanics. Instead, our approach to quantum mechanics 
is modelled on a statement that is close in spirit to the equivalence 
principle of general relativity, thus bearing a strong resemblance with the 
equivalence principle of quantum mechanics formulated by Faraggi--Matone. 

\end{abstract}

\section{Introduction}\label{intro}

\subsection{Motivation}\label{moti}

Quantisation may be understood as a prescription to construct a quantum theory from a given 
classical theory. As such, it is far from being unique. Beyond canonical quantisation 
and Feynman's path--integral, a number of  different, often complementary approaches 
to quantisation are known, each one of them exploiting different aspects of the underlying classical
theory. To mention just a few, the geometric quantisation of Kirillov, Kostant and Souriau relies
heavily on the theory of group representations \cite{KIRILLOV, KOSTANT, SOURIAU, WOODHOUSE}.
Berezin's quantisation can be applied to classical systems whose  phase space is a homogeneous
K\"ahler manifold \cite{BEREZIN}. The approach of Kontsevich \cite{KONTSEVICH}
is based on the deformation quantisation of Poisson manifolds \cite{FELDER, SCHLICHENMAIER}, 
bearing some resemblance with noncommutative geometry. 

The deep link existing between classical and quantum mechanics has of course been known 
for long. Perhaps its simplest manifestation is that of coherent states \cite{COHST}. 
More recent is the notion that {\it not all quantum theories arise from quantising 
a classical system}. Furthermore, a given quantum model 
{\it may possess more than just one classical limit}.
These ideas find strong evidence in string duality and M--theory \cite{SCHWARZ, VAFA}.
 
It therefore seems natural to try an approach to quantum mechanics that is not based, 
at least primarily, on the the quantisation of a given classical dynamics. 
In such an approach one would {\it not} take a classical theory as a starting point. 
Rather, quantum mechanics itself would be more fundamental, in that its classical 
limit or limits (possibly more than one) would follow from a parent quantum theory.

In order to carry out this programme one can envisage two different aproaches 
to first quantisation, one technical, the other  conceptual. A technical approach 
has been presented in \cite{NOS}, in connection with the quantum--mechanical 
implementation of an S--duality symmetry. In this paper we would like to take 
a more conceptual viewpoint instead.

\subsection{Summary}\label{suma}

The general purpose of this paper is to analyse some properties that quantum mechanics 
must satisfy, if it is {\it not} to be formulated as a quantisation of a given classical 
mechanics. We will formulate a statement, close in spirit to the 
equivalence principle of general relativity, that could well provide a 
starting point for a reformulation of quantum mechanics such as that claimed 
by Vafa \cite{VAFA}. Our formalism may be understood as a certain limit of Berezin's 
quantisation \cite{BEREZIN}. 
The latter relies on the metric properties of classical phase space ${\cal M}$, 
whenever ${\cal M}$ is a homogeneous K\"ahler manifold. In Berezin's method, 
quantum numbers arise naturally from the metric on ${\cal M}$. The semiclassical regime 
is then identified with the regime of large quantum numbers. Our method may be regarded 
as the topological limit of Berezin's quantisation, in that the metric dependence 
has been removed. As a consequence of this topological nature our quantum mechanics 
exhibits some added features. Quantum numbers are not originally present in our prescription; 
they appear only after a vacuum has been chosen, and even then they are local in nature, 
instead of global. Hence our procedure may be thought of as a manifestly non--perturbative 
formulation of quantum mechanics, in that we take no classical phase space and no Poisson brackets 
as our starting point, {\it i.e.}, we do not deform a classical theory into 
its quantum counterpart, as in deformation quantisation.

\subsection{Outline}\label{outl}

This paper is organised as follows. Section \ref{berezin} sets the scene by 
giving a quick review of Berezin's quantisation, starting from the metric 
on certain classical phase spaces. Section \ref{epfm} summarises the 
equivalence principle of Faraggi--Matone \cite{MATONE}, as a preparation for our own 
starting point. The latter is presented under the
form of a statement in section \ref{statement}. Its physical implications 
are analysed and discussed in detail in section \ref{discussion}.

\section{Berezin's Quantisation}\label{berezin}

Below we sketch the construction of the Hilbert space of states from the metric 
on some relevant homogeneous K\"ahler manifolds \cite{BEREZIN}.

\subsection{The Complex Plane}\label{complane}

Coherent states $|z\rangle$ of standard quantum mechanics on ${\bf R}$  are defined as the 
eigenstates of the annihilation operator $a=(Q+{\rm i}P)/\sqrt{2\hbar}$.
The Hilbert space of states is the Fock-Bargmann space ${\cal 
F}_{\hbar}({\bf C})$ of entire analytic functions
$\psi(z)$ on ${\bf C}$ with finite norm with respect to the scalar product
\begin{equation}
\langle\psi_1|\psi_2\rangle =\int_{\bf C}{\rm d}\mu(z,\bar z)\,{\rm exp}(-z\bar
z)\,{\overline\psi_1(z)}\psi_2(z),
\label{fbscalar}
\end{equation} 
the measure being given by ${\rm d}\mu(z,\bar z)=\pi^{-1} {\rm d} 
z\wedge {\rm d}\bar z$. The latter follows from
the K\"ahler form $\omega(z,\bar z)={\rm d} z\wedge {\rm d}\bar z$ on ${\bf C}$. 
The argument of the exponential equals the K\"ahler potential 
$K_{\bf C}(z,\bar z)=z\bar z$ for the flat metric $g_{z\bar z}=1$. 
There is a natural isomorphism between the oscillator states $|n\rangle$ 
and the basis states $z^n/\sqrt{n!}$, $n=0,1,\ldots$ of the Fock-Bargmann space.
The semiclassical limit corresponds to letting $n\to\infty$. 

\subsection{The Riemann Sphere}\label{riesphere}

On the Riemann sphere $S^2$, the K\"ahler potential $K_{S^2}(z,\bar z)={\rm log}\,(1+|z|^2)$
produces an integration measure ${\rm d}\mu(z,\bar z)=(2\pi {\rm i})^{-1}{\rm d} 
z\wedge {\rm d}\bar z/ (1+|z|^2)^2$.
The Hilbert space of states now becomes the space ${\cal F}_{\hbar}(S^2)$ 
of holomorphic functions on
$S^2$ with finite norm, the scalar product being
\begin{equation}
\langle\psi_1|\psi_2\rangle =\Big({1\over \hbar} +1\Big)\int_{S^2}{\rm d}\mu(z,\bar
z)\,(1+|z|^2)^{-1/\hbar}\,{\overline
\psi_1(z)}\,\psi_2(z).
\label{sphscalar}
\end{equation} 
It turns out that $\hbar^{-1}$ must be an integer. For $\psi$ to have finite
norm, it must be a polynomial of degree less than $\hbar^{-1}$. In fact, setting $\hbar^{-1}=2j+2$, 
${\cal F}_{\hbar}(S^2)$ is the representation space for the spin-$j$ representation of $SU(2)$. The
semiclassical regime now corresponds to $j\to\infty$. Coherent states $|u\rangle$ are parametrised
by points $u$ in the quotient space $S^2=SU(2)/U(1)$. 

\subsection{The Lobachevsky Plane}\label{lobachevsky}

Consider the Lobachevsky plane modelled as the unit disc
$D=\{z\in {\bf C}:|z|<1\}$. From the K\"ahler potential $K_{D}(z,\bar z)=-{\rm log}\,(1-|z|^2)$ 
one derives an integration measure ${\rm d}\mu(z,\bar z)=
(2\pi {\rm i})^{-1}{\rm d} z\wedge {\rm d}\bar z/ (1-|z|^2)^2$.  
With respect to the scalar product
\begin{equation}
\langle\psi_1|\psi_2\rangle =\Big({1\over \hbar} -1\Big)\int_{D}{\rm d}\mu(z,\bar
z)\,(1-|z|^2)^{1/\hbar}\,{\overline\psi_1(z)}\,\psi_2(z),
\label{lobscalar}
\end{equation} 
the space ${\cal F}_{\hbar}(D)$ of analytic functions on $D$ with finite norm defines a  Hilbert
space of states. Setting $k-1=(2\hbar)^{-1}$,  ${\cal F}_{\hbar}(D)$ becomes the
representation space for the discrete series of the group $SU(1,1)$ of isometries of $D$, {\it
i.e.}, the space of weight-$k$ modular forms. Large values of $k$ correspond to the semiclassical
limit of this quantum mechanics. Coherent states $|w\rangle$ are parametrised by points $w$ in the
quotient space $SU(1,1)/U(1)$.

\subsection{Complex Homogeneous K\"ahler Manifolds}\label{complex}

Let $z^j$, $\bar z^k$, $j, k= 1,\ldots, n$, be  local coordinates on a
complex homogenous K\"ahler manifold ${\cal M}$, and let $K_{\cal M}(z^j,\bar z^k)$ be a K\"ahler
potential for the metric ${\rm d} s^2=g_{j\bar k}\,{\rm d} z^j{\rm d}\bar z^k$. The K\"ahler form
$\omega=g_{j\bar k}\,{\rm d} z^j\wedge {\rm d}\bar z^k$ gives rise to an integration measure
${\rm d}\mu(z,\bar z)$, 
\begin{equation} 
{\rm d}\mu(z,\bar z)=\omega^n={\rm det}\,(g_{j\bar k})\,\prod_{l=1}^n{{\rm d} z^l\wedge {\rm d}\bar
z^l\over 2\pi {\rm i}}.
\label{bermeasure}
\end{equation} 
The Hilbert space of states is the space ${\cal F}_{\hbar}({\cal M})$ of analytic
functions on
${\cal M}$ with finite norm, the scalar product being
\begin{equation}
\langle\psi_1|\psi_2\rangle =c(\hbar)\,\int_{\cal M}{\rm d}\mu(z,\bar z)\,{\rm
exp}(-\hbar^{-1}K_{\cal M}(z,\bar z))\,{\overline
\psi_1(z)}\psi_2(z),
\label{berscalar}
\end{equation}
and $c(\hbar)$ a normalisation factor.  Let  $G$ denote the Lie group of motions of ${\cal M}$, 
and assume $K_{\cal M}(z,\bar z)$ is invariant under $G$. Setting $\hbar=k^{-1}$, the family
of Hilbert spaces ${\cal F}_{\hbar}({\cal M})$ provides a discrete series of projectively unitary
representations of $G$. The homogeneity of ${\cal M}$ is used  to prove that the correspondence
principle is satisfied in the limit $k\to\infty$. Furthermore, let $G'\subset G$ be a maximal
isotropy subgroup of the vacuum state $|0\rangle $. Then coherent states $|\zeta\rangle $ are
parametrised by points $\zeta$ in the coset space $G/G'$.

\section{The Equivalence Principle of Faraggi--Matone}\label{epfm}

In \cite{MATONE} an entirely new presentation of quantum mechanics has been given, 
starting from the so--called {\it equivalence principle of quantum mechanics}. 
In plain words, the philosophy underlying this approach could be summarised 
as follows. The classical Hamilton--Jacobi technique is based on transforming 
an arbitrary dynamical system, by means of coordinate changes, into a freely--evolving 
system subject to no interactions. The requirement that this equivalence also hold 
in the case when the conjugate variables are considered as dependent 
leads to a quantum analogue of the Hamilton--Jacobi equation corresponding 
to the one previously assumed by Floyd \cite{FLOYD}. See also the book by 
Carroll \cite{CARROLL}.

A surprising new feature of the Faraggi--Matone approach is that the 
quantum analogue of the Hamiltonian characteristic function is quite 
different from the usual one in the literature. This solves Einstein's 
criticism to Bohm's approach (see \cite{HOLLAND} p. 243). As a result, 
the quantum potential is never trivial. The latter has been used in 
\cite{MMATONE} to derive the gravitational interaction. 
This suggests that gravitation would have a quantum origin, 
so that physical interactions would {\it not} have to be introduced by hand, 
through the consideration of a potential. Rather, they would {\it follow} 
from consistency requirements. In particular, the quantum potential term 
that it is customary to neglect in the semiclassical limit plays a decisive 
role in determining the interaction, precisely due to the observation that 
it should not be neglected.

\section{The Opening Statement}\label{statement}

The equivalence principle of general relativity states that \cite{WEINBERG}

{\it At every spacetime point in an arbitrary gravitational field, it is 
possible to choose a locally inertial coordinate system such that, within 
a sufficiently small region of the point in question, the laws of motion 
take the same form as in unaccelerated Cartesian coordinate systems in the 
absence of gravitation.}

Let us now make the opening statement that

{\it Given any quantum system, there always exists a coordinate transformation 
that transforms the system into the semiclassical regime, i.e., into a system that 
can be studied by means of a perturbation series in powers of $\hbar$ 
around a certain local vacuum.}

One can perceive a conceptual analogy with the equivalence principle of 
Faraggi--Matone in the use of coordinate transformations in order to trivialise 
a given system. In our context, however, {\it trivialisation} does not mean cancellation 
of the interaction term, as in the Hamilton--Jacobi approach. Rather, 
it refers to the choice of a vacuum around which to perform a perturbative 
expansion in powers of $\hbar$. As we will see presently, this is 
equivalent to eliminating the metric, thus rendering quantum mechanics 
metrically trivial.

\section{Physical Discussion}\label{discussion}

\subsection{The Choice of a Vacuum}\label{vuoto}

The opening statement above instructs us to choose a local vacuum.
Under the {\it choice of a vacuum} we understand a specific set of coordinates
around which to perform an expansion in powers of $\hbar$. This choice of a vacuum is 
{\it local in nature}, in that it is linked to a specific choice of coordinates. 
It breaks the group of allowed coordinate transformations to a (possibly 
discrete) subgroup, leaving behind a (possibly discrete) duality symmetry of the quantum 
theory. Call $q$ the local coordinate corresponding to the vacuum in 
question, and $Q$ its quantum operator. The corresponding local momentum $P$ 
satisfies the usual Heisenberg algebra with $Q$. However, as our starting point 
we have no classical phase space at all, and no Poisson brackets to quantise 
into commutators. This may be regarded as a manifestly non--perturbative formulation 
of quantum mechanics, such as that claimed in \cite{VAFA}. 

\subsection{Quantum Numbers vs. a Topological Quantum Mechanics}\label{quantumnumbers}

Berezin's quantisation relied heavily on the metric properties of classical phase space. 
The semiclassical limit could be defined as the regime of large quantum numbers. 
The very existence of quantum numbers was a consequence of the metric structure.
If quantum mechanics is {\it not} to be formulated as a quantisation of a 
given classical mechanics, then we had better do away with global quantum 
numbers, {\it i.e.}, with the metric. Metric--free theories usually go by 
the name of topological theories. Hence  our quantum mechanics will be a
{\it topological quantum mechanics}, {\it i.e.} free of {\it global} quantum numbers.  
Locally, of course, quantum numbers do appear, but only after the choice of a {\it local} vacuum. 

\subsection{Classical vs. Quantum} \label{versus}

A feature of this approach is the following. After the choice of a local 
vacuum to expand around, the local quantum numbers one obtains describe
excitations around the local vacuum chosen. Hence what appears to be a 
semiclassical excitation to a local observer may well turn out to be a 
highly quantum phenomenon, when described from the viewpoint of a different 
local vacuum.

The logic could be summarised as follows: 
\begin{itemize}
\item[1.]{the fact that this quantum mechanics is topological implies the  absence of
a metric;}
\item[2.]{the absence of a metric implies the absence of global quantum numbers;}
\item[3.]{the absence of global quantum numbers implies the impossibility of globally 
defining a semiclassical regime. The latter exists only locally.}
\end{itemize}

\subsection{An example}

In order to explore the quantum dynamics of a given system we need a 
knowledge of the Hilbert space of states and of the operators acting on it.
Our opening statement does not determine them, so they can only be specified
on a case--by--case basis, by imposing consistency with the symmetry requirements 
of the particular problem at hand. The same applies to the group of coordinate 
transformations that our opening statement refers to. We  illustrate these points
with the particular case of a point particle on the real line 
analysed in \cite{NOS}.

The basic requirement imposed in \cite{NOS} is that an S--duality transformation 
(modelled on ${\bf Z}_2$) exist between the semiclassical and the strong quantum regimes. 
This determines the group of coordinate transformations to be ${\rm SL}(2,{\bf R})$.
The Hilbert space of a point particle on the real line turns out 
to be $L^2(0,\infty)$, which strictly contains $L^2({\bf R})$. States belonging 
to $L^2(0,\infty)$ but not to $L^2({\bf R})$ may be interpreted as nonperturbative 
states, that cannot be reached by the standard perturbative approach. The usual 
Hilbert space $L^2({\bf R})$ only emerges after the choice of a vacuum.
There are two possible vacua, $|0_z\rangle$ and $|0_{\tilde z}\rangle$,
correponding to the coordinates $z$ and $\tilde z=-z^{-1}$ 
on the upper half--plane ${\bf H}$. The ${\rm SL}(2,{\bf R})$ symmetry acting 
homogeneously on ${\bf H}$ breaks down to the affine group, times ${\bf Z}_2$.
The latter implements the desired S-duality symmetry exchanging the semiclassical 
and the strong quantum regimes of one single quantum theory. The coherent states 
constructed around $|0_z\rangle$ are not coherent from the viewpoint of 
$|0_{\tilde z}\rangle$. This model does not allow for globally defined coherent 
states such as those of section \ref{berezin}, thus providing an explicit example 
of the general procedure presented above. Thus the quantum mechanics developed in \cite{NOS} 
is that of affine variables on the real line. (Incidentally, affine symmetry is well--suited 
to quantum gravity \cite{ISHAMKAKAS, AFFKLAUDER, 1DIMAFF, WATSON}). 

{\bf Acknowledgments}

It is a very great pleasure to thank Marco Matone for many conversations that made 
the preparation of this work much more enjoyable. The author would also like to thank 
Diego Bellisai, Kurt Lechner, Pieralberto Marchetti, Paolo Pasti, Dmitri Sorokin and 
Mario Tonin. This work has been supported by a Fellowship from Istituto Nazionale di 
Fisica Nucleare (Italy).

\end{document}